\documentclass{l4dc2027}
\usepackage{times,booktabs,placeins}
\makeatletter
\let\arxivtitlepagestyle\ps@jmlrtps
\renewcommand{\ps@jmlrtps}{%
  \arxivtitlepagestyle
  \let\@oddhead\@empty
  \let\@evenhead\@empty
}
\makeatother
\title[Enforcing Ellipsoidal Constraints with a Polynomial Safety Layer]{Enforcing Ellipsoidal Constraints in Neural Networks with a Polynomial Safety Layer}
\hypersetup{pdftitle={Enforcing Ellipsoidal Constraints in Neural Networks with a Polynomial Safety Layer}}
\newif\ifprintauthors
\printauthorstrue
\ifprintauthors
\author[Findeisen et al.]{%
\Name{Rolf Findeisen}$^{1}$ \Email{rolf.findeisen@iat.tu-darmstadt.de}\\
\Name{Sebastian Hirt}$^{1}$ \Email{sebastian.hirt@iat.tu-darmstadt.de}\\
\Name{Hoang Hai Nguyen}$^{1}$ \Email{hoang.nguyen@iat.tu-darmstadt.de}\\
\Name{Grace Li Zhang}$^{2}$ \Email{grace.zhang@tu-darmstadt.de}\\
\Name{Hendrik Alsmeier}$^{1}$ \Email{hendrik.alsmeier@iat.tu-darmstadt.de}\\
\addr $^{1}$Control and Cyber-Physical Systems Laboratory, 
\addr $^{2}$Hardware for Artificial Intelligence \\
\addr Technical University of Darmstadt, Germany}
\else
\author{}

\makeatletter
\renewcommand*{\@titlefoot}{}
\makeatother
\fi
\begin{document}
\raggedbottom\maketitle
\begin{abstract} 
Often the output of a learned mapping must satisfy prescribed constraints. In the control of autonomous systems, for example, the chosen action must keep the system within a collision-free region. We consider the case in which a proposed output is corrected by projection onto a safety ellipsoid. Computing this projection amounts to solving a quadratically constrained quadratic program (QCQP). We construct a neural-network safety layer by unrolling a feasibility-preserving polynomial penalty iteration. The unrolled structure allows the loss gradient to propagate from the feasible output through the safety layer to the network parameters during training. We show how to choose the layer parameters so that every iterate is feasible and give a computable bound on its distance to the exact projection at any prescribed depth. We illustrate the method by training a LiDAR-based policy to control an unmanned aerial vehicle. A privileged model predictive control expert uses the map and the measured vehicle state to generate the training data. In the simulated narrow mazes, the filtered policy avoided wall contact at every tested stage of training, including at initialization, when every unfiltered flight hit a wall. A representative build of the learned policy and its safety layer for an Arm Cortex-M4F microcontroller occupies about 36~KiB of flash, with a RAM budget below 5~KiB.
\end{abstract}
\begin{keywords}
safety filters, hard output constraints, quadratically constrained quadratic programs,
algorithm unrolling, imitation learning, model predictive control, control barrier functions
\end{keywords}
\section{Introduction}
In many applications, the output of a learned mapping must satisfy a hard constraint at each evaluation. For example, a network that allocates assets must respect a prescribed risk budget. For the control of an autonomous system, the command returned by a learned policy must belong to a feasible set to ensure safety and collision avoidance. Even if an expert supplies only feasible commands during training, the policy can propose an infeasible command~\citep{drgona25}. Safety filters can correct the proposal before it is applied~\citep{hsu2024safety,wabersich2023filters,drgona25}. Embedding the safety layer in the policy allows training to account for the corrected action~\citep{dalal2018safe}.

\begin{figure}[!t]\centering
\includegraphics[width=.87\linewidth,trim=0bp 7bp 0bp 5bp,clip]{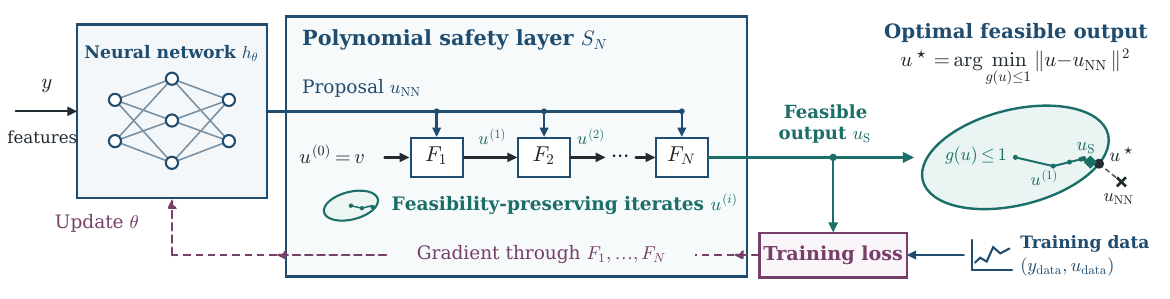}
\caption{The neural network \(h_\theta\) proposes \(u_{\mathrm{NN}}\). The safety layer starts at the feasible point \(u^{(0)}=v\). Each block \(F_\ell\) performs one polynomial penalty step and returns a feasible iterate. The diamond marks the filtered output \(u_{\mathrm S}\), and the black dot marks the exact projection \(u^\star\). Pink shows the learning process. During training, \(y=y_{\mathrm{data}}\) and the loss compares \(u_{\mathrm S}\) with the paired target \(u_{\mathrm{data}}\).
}\label{fig:arch}
\end{figure}

We consider learned mappings whose outputs must satisfy quadratic constraints. When the feasible outputs form an ellipsoid, finding the smallest correction to a proposal amounts to solving a quadratically constrained quadratic program (QCQP). Such constraints arise, for example, from discrete-time barrier conditions with a quadratic barrier for systems whose next state is affine in the input~\citep{agrawal2017dcbf}. To account for the correction during training, we implement the safety layer as a sequence of neural network blocks at the output of the learned mapping (Figure~\ref{fig:arch}). Each block performs one step of the polynomial penalty iteration. Starting from a feasible output, the iteration preserves feasibility under an explicit rule for choosing the coefficients. A scalar recursion bounds the distance from the filtered output to the exact projection at any chosen depth.

A safety filter defined by a convex optimization problem can be incorporated into training by differentiating through its solution~\citep{agrawal2019cvxpylayers}. For example, a neural controller can enforce control barrier function constraints through a differentiable quadratic program~\citep{xiao2023barriernet}. Other approaches parameterize feasible predictions directly~\citep{rayen2023} or incorporate constraint corrections into training~\citep{donti2021dc3}.  Unrolling assigns one iteration of an optimization algorithm to each network block~\citep{gregor2010lista}. In our construction each polynomial iteration is implemented exactly as a feedforward neural network block for the given coefficients. At any prescribed depth the layer gives a bound on the projection error. With the layer parameters chosen according to our rule, every block returns a feasible output for any network proposal during training, adaptation and deployment.

Section~\ref{sec:certificate} gives the construction and the finite-depth certificate. Section~\ref{sec:navigation} applies it to a LiDAR-based policy trained to imitate a privileged model predictive controller for a drone. We examine the commands proposed during learning and the actions returned by the filter for the drone.
\section{Safety layer from polynomial iterations with feasibility guarantees}\label{sec:certificate}
Let \(y\) denote the features supplied to the neural network \(h_\theta\) with trainable weights \(\theta\).  \(u_{\mathrm{NN}}=h_\theta(y)\) is the proposal provided by the network. The feasible set \(\mathcal C\), which for example describes the safe set with respect to collision avoidance, is given by the center \(v\) and the positive definite shape matrix \(M\) by \(\mathcal C=\{u:g(u)\leq 1\}\), where \(g(u)=(u-v)^\top M(u-v)\). The feasible output closest to the network proposal is
\begin{equation}
 u^\star=\operatorname*{arg\,min}_{u\in\mathcal C}
 f(u),\qquad f(u)=\tfrac12\|u-u_{\mathrm{NN}}\|_2^2 .
 \label{eq:projection}
\end{equation}
This QCQP has a unique solution and returns the proposal itself whenever it is feasible. For a proposal outside the ellipsoid, the exact projection \(u^\star\) lies on its boundary. An approximation can therefore be arbitrarily close to \(u^\star\) and still lie outside. We require each iteration block to return a point in \(\mathcal C\). Note that any constraint \(u^\top Q u+2b^\top u+c\leq0\) with \(Q\succ0\) and nonempty interior takes this form with \(v=-Q^{-1}b\), \(\rho=b^\top Q^{-1}b-c\), and \(M=Q/\rho\). A positive definite weight on the correction can likewise be absorbed by a change of coordinates.

\subsection{Polynomial iteration as a neural network for the projection}
Our main contribution is to construct a neural network safety layer from the feasibility-preserving polynomial penalty iteration developed for encrypted QCQP optimization~\citep{schlor2025}. The exact realization integrates the safety mechanism into a feedforward architecture with a prescribed depth and feasible block outputs. Each of its \(N\) blocks implements one iteration exactly using affine maps and a square activation function \(\sigma(\tau)=\tau^2\). Policy and safety layer thus form a single feedforward network of fixed depth, with the safety layer using only additions and multiplications for given coefficients. This makes the combined network a candidate for implementation as one network with fixed latency on dedicated hardware, for example on FPGAs or as logic circuits~\citetext{\citealp{Xu2024LogicDesign}; \citealp{Lucia2018FPGA}}. From a feasible initial iterate, every iteration block returns a feasible output. During training, the loss at the chosen block output can be differentiated through the blocks to update the preceding network. Furthermore, for the projection problem in~\eqref{eq:projection}, we give an explicit choice of the penalty weight \(m\) and bound the distance to the exact projection at each depth. 

Each polynomial iteration block \(F_{\ell+1}\) takes the current iterate \(u^{(\ell)}\) and the network proposal \(u_{\mathrm{NN}}\) as inputs and returns the next iterate \(u^{(\ell+1)}\). Starting from \(u^{(0)}=v\), each block takes one gradient step of length \(\gamma_\ell\) on the penalized objective \(J_{k_\ell}=f+(m/k_\ell)g^{k_\ell}\). Here \(m\ge0\) is the penalty weight shared by all blocks during one evaluation, and \(k_\ell\ge1\) is the integer exponent chosen for block \(F_{\ell+1}\). All blocks use the same ellipsoid data \((v,M)\). For \(\ell=0,\ldots,N-1\) this gives
\begin{equation}
 u^{(\ell+1)}=u^{(\ell)}-\gamma_\ell
 \big[u^{(\ell)}-u_{\mathrm{NN}}+2m\,g(u^{(\ell)})^{k_\ell-1}M(u^{(\ell)}-v)\big].
 \label{eq:filter}
\end{equation}
For a given exponent \(k_\ell\), the ellipsoid data \((v,M)\), penalty weight \(m\) and step size \(\gamma_\ell\) determine the coefficients of the polynomial in \(u^{(\ell)}\) and \(u_{\mathrm{NN}}\). We therefore call \(F_{\ell+1}\) a polynomial iteration block. To implement this update, the block first computes \(d=u^{(\ell)}-v\) and \(Md\) by affine maps. Two neurons with square activation \(\sigma\) suffice to multiply two scalars. They take \(a+b\) and \(a-b\) as inputs, and their outputs give \(ab=[\sigma(a+b)-\sigma(a-b)]/4\). Repeating this construction and summing the products gives \(g(u^{(\ell)})=d^\top Md\). Further products give the vector \(g(u^{(\ell)})^{k_\ell-1}Md\). An affine output combines this vector with \(u^{(\ell)}\) and \(u_{\mathrm{NN}}\) to give exactly the iterate in~\eqref{eq:filter}. After \(N\) such blocks the layer returns \(u_{\mathrm S}=S_N(u_{\mathrm{NN}};v,M)=u^{(N)}\). For a given ellipsoid, the complete layer is polynomial in the proposal if its coefficients are chosen in advance for a bounded range of proposals. In the navigation example the current proposal determines the penalty weight and step sizes before the blocks are evaluated.

\subsection{Feasibility, projection error and convergence}
The safety layer must return a feasible output at any prescribed depth. The following theorem guarantees this property for suitable choices of \(m\), the weight of the penalty term \(g^{k_\ell}/k_\ell\) in the objective \(J_{k_\ell}\), and of the block step sizes \(\gamma_\ell\). 
\begin{theorem}[Feasibility and projection error]\label{thm:exit}
Let \(\lambda_-,\lambda_+>0\) be the smallest and largest eigenvalues of \(M\). Initialize \(u^{(0)}=v\), choose an integer \(k_\ell\ge1\) for each step, and keep the penalty weight \(m\) fixed throughout one network evaluation. Let \(m\) and the step sizes \(\gamma_\ell\) satisfy
\begin{equation}
 m\ge\frac{\max(0,\sqrt{g(u_{\mathrm{NN}})}-1)}{2\lambda_-},\qquad
 0<\gamma_\ell\le [1+(4k_\ell-2)m\lambda_+]^{-1}.\label{eq:scaling}
\end{equation}
With \(E_0=1/\lambda_-\) and \(q_\ell=(1-\gamma_\ell)/(1+\gamma_\ell)\), define
\begin{equation}
 E_{\ell+1}=q_\ell E_\ell+(1-q_\ell)m/k_\ell .\label{eq:certificate}
\end{equation}
Then \(g(u^{(N)})\le1\) and \(\|u^{(N)}-u^\star\|_2\le\sqrt{E_N}\) for every \(N\ge0\).
\end{theorem}
\begin{proof}
Use \(u,u^+\) for successive iterates. A change of variables \(w=M^{1/2}(u-v)\) maps the ellipsoid to the unit ball, so it suffices to show that \(r=\|w\|\le1\) implies \(\|w^+\|\le1\). With \(\tilde d=M^{1/2}(u_{\mathrm{NN}}-v)\), the update becomes \(w^+=Aw+\gamma\tilde d\), where \(A=(1-\gamma)I-2\gamma m r^{2k-2}M\succeq0\) by~\eqref{eq:scaling}. Its largest eigenvalue gives \(\|w^+\|\le(1-\gamma)r-2\gamma m\lambda_-r^{2k-1}+\gamma\|\tilde d\|\). By~\eqref{eq:scaling} this bound is nondecreasing in \(r\) on \([0,1]\) and at most one at \(r=1\), since \(\|\tilde d\|=\sqrt{g(u_{\mathrm{NN}})}\). Induction from \(w^{(0)}=0\) gives feasibility. We bound the squared error \(D=\|u-u^\star\|^2\) without computing \(u^\star\). Write \(D^+=\|u^+-u^\star\|^2\). On the ellipsoid, \(\|\nabla g\|^2\le4\lambda_+g\) gives \(I\preceq\nabla^2 J_k\preceq[1+(4k-2)m\lambda_+]I\preceq I/\gamma\). For \(\Delta J=J_k(u^+)-J_k(u^\star)\), the quadratic upper bound on the segment \([u,u^+]\) and strong convexity at \(u\) give \(2\gamma\Delta J\le(1-\gamma)D-D^+\), using \(u^+-u=-\gamma\nabla J_k(u)\). Since \(u^+\) is feasible, the optimality condition for the Euclidean projection \(u^\star\) gives \(f(u^+)-f(u^\star)\ge D^+/2\). Since \(0\le (m/k)g^k\le m/k\) on the ellipsoid, \(\Delta J\ge D^+/2-m/k\). Thus \(D^+\le qD+(1-q)m/k\). The initial bound \(\|v-u^\star\|^2\le E_0\) and induction yield \(\|u^{(N)}-u^\star\|^2\le E_N\) by~\eqref{eq:certificate}.
\end{proof}

The feasibility guarantee is independent of the architecture of the preceding network and applies equally to proposals generated by convolutional networks or transformers, provided the conditions of Theorem~\ref{thm:exit} hold.

With a linearly increasing penalty exponent and the largest permitted step sizes, the block outputs converge to the exact projection while remaining feasible at every depth. The following corollary quantifies this convergence.
\begin{corollary}[Decay of the projection error bound]\label{cor:rate}
Take \(k_\ell=\ell+1\) and the largest step in~\eqref{eq:scaling}. For fixed \(\kappa=2m\lambda_+>0\), \(E_N\) is \(O(N^{-1})\) if \(\kappa<1\), \(O(N^{-1}\log N)\) if \(\kappa=1\), and \(O(N^{-1/\kappa})\) if \(\kappa>1\).
\end{corollary}
\begin{proof}
Set \(w_j=1-q_{j-1}=[1+\kappa(j-\tfrac12)]^{-1}\). Since \(1-w_j\le e^{-w_j}\), integral comparison gives \(\prod_{j=i+1}^N(1-w_j)\le[(1+\kappa(i+\tfrac12))/(1+\kappa(N+\tfrac12))]^{1/\kappa}\) for \(0\le i\le N\). Expanding~\eqref{eq:certificate} with this bound and \(w_i=O(i^{-1})\) yields \(E_N=O\bigl(N^{-1/\kappa}(1+\sum_{i=1}^N i^{1/\kappa-2})\bigr)\). Summing the powers of \(i\) gives the three rates.
\end{proof}

\begin{remark}[Learning through the safety layer]
The training dataset contains pairs \((y_{\mathrm{data}},u_{\mathrm{data}})\) of input features and target outputs. For each pair, the network proposes \(h_\theta(y_{\mathrm{data}})\), and the loss compares the filtered output \(u_{\mathrm S}\) with the target \(u_{\mathrm{data}}\). Backpropagation through the \(N\) blocks gives the gradient with respect to \(\theta\). The penalty weight \(m\) and step sizes \(\gamma_\ell\) can also be chosen as functions of the network proposal, as in the navigation example in Section~\ref{sec:navigation}. Backpropagation then includes the dependence of both \(m\) and \(\gamma_\ell\) on the proposal. When the filtered output is locally insensitive to the proposal, the gradient reaching the network can vanish. In that example, an auxiliary loss compares the proposal directly with the expert action.
\end{remark}
\section{Learning a LiDAR-based drone navigation policy from a privileged MPC expert}\label{sec:navigation}
\begin{figure}[!t]\centering
\includegraphics[width=.88\linewidth,trim=0bp 7bp 0bp 1bp,clip]{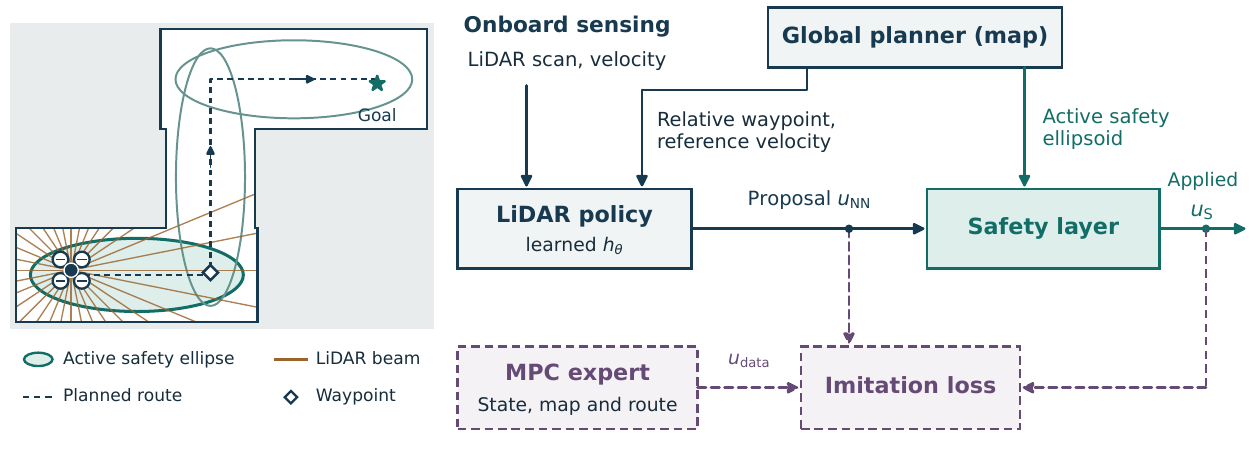}
\caption{The learned LiDAR policy proposes an acceleration from sensor measurements and route references. The safety layer filters the proposal before it is applied. The MPC supplies the training targets.}\label{fig:lidar-principle}
\end{figure}

We consider an indoor drone that must reach a goal through a maze without hitting the walls (Figure~\ref{fig:lidar-principle}). A global planner uses the map to provide a feasible route and waypoints for the local controller~\citep{richter2016,koegel25}. We train a LiDAR-based policy by imitating an expert with access to additional information~\citep{chen2020}. Here the expert is a model predictive controller (MPC) that uses the measured state and the mapped corridor ahead. Its first acceleration command serves as the training target. The policy uses local range measurements for navigation~\citep{pfeiffer2017}. At each update it proposes a planar acceleration from the latest scan, the measured velocity, the relative waypoint and the reference velocity.

Collision-free demonstrations alone do not guarantee that the learned policy satisfies the safety constraints~\citep{hsu2024safety,drgona25}. At a bend, for example, an imitation error can cause the drone to brake too late and hit the outer wall (Figure~\ref{fig:paper-intervention}). The planner covers the route with overlapping cells. Each cell is an ellipsoid \(\{x : V_i(x)\le 1\}\) in the joint state \(x=(p,w)\) of position and velocity, with a quadratic function \(V_i\) that is positive definite about the cell center. The planner selects one active cell, also where two cells overlap. With the planar acceleration \(u\in\mathbb{R}^2\) held constant over one sampling interval \(T\), the nominal next state \(x^+(u)\) is affine in \(u\). For the cells used here, \(V_i(x^+(u)) \le (1-\beta)V_i(x)+\beta\), with \(0<\beta<1\), therefore defines a nonempty ellipse of accelerations that keep the next state in the cell when \(V_i(x)\le1\). With \(h_i = 1-V_i\), this is the discrete-time barrier condition \(h_i(x^+(u)) \ge (1-\beta)h_i(x)\). Each cell also limits the velocity so that the drone, continuing at its current velocity for half a sampling interval, would still be inside the cell's position ellipse. Together with the current and the next position, this point \(p+Tw/2\) spans a triangle that contains the flight segment \(p+tw+\tfrac12 t^2u\), \(0\le t\le T\). Since the ellipse is convex, the whole nominal segment stays in mapped free space. The planner switches to the next cell only once the measured position and velocity satisfy its entry condition.

\begin{table}[!b]\centering\small
\caption{Goal arrivals / wall contacts in 108 narrow-maze flights per column and training budget. The remaining flights timed out.}\label{tab:paper-flights}
\begin{tabular}{@{}lcccc@{}}\toprule
& \multicolumn{2}{c}{Trained with filter} & \multicolumn{2}{c}{Trained without filter} \\
\cmidrule(lr){2-3}\cmidrule(l){4-5}
Examples & Filter active & Bypassed & Filter added & No filter \\\midrule
10,000 & 102 / 0 & 78 / 6 & 60 / 0 & 90 / 0 \\
20,000 & 90 / 0 & 54 / 0 & 102 / 0 & 90 / 0 \\
Full dataset & 96 / 0 & 72 / 0 & 102 / 0 & 102 / 0 \\\bottomrule
\end{tabular}
\end{table}

\subsection{MPC expert and policy training}
 To train the LiDAR policy, we pair each observation with the acceleration commanded by the expert in closed-loop simulation.
\paragraph{Privileged MPC expert.}
At each sampling instant the expert uses the measured state to solve a constrained tracking problem over a receding horizon~\citep{rawlings:mayne:diehl:2020,findeisen03c}. It predicts planar motion with two acceleration inputs over \(3.5\,\mathrm{s}\) at a sampling interval of \(0.1\,\mathrm{s}\). It tracks the planner's waypoints and reference velocities, constrains each predicted flight segment to its assigned corridor cell and checks membership in both cells at a handover. The first acceleration must also lie in the same safety ellipse used by the learned policy. The expert applies this acceleration for one sampling interval and then replans.

\paragraph{Training data.} We collected about 83,000 expert actions on 296 layouts with narrow and wide corridors. We verified each accepted plan over its full prediction horizon.

\paragraph{LiDAR policy.} The navigation example uses a compact multilayer perceptron whose embedded memory requirements are reported in Section~\ref{sec:computational-effort}. The policy receives 32 LiDAR ranges, the measured velocity, the relative waypoint and the reference velocity. Two hidden layers of 64 tanh units produce the two acceleration components. The safety layer uses ten polynomial iteration blocks with exponents \(k_\ell=\ell+1\) for \(\ell=0,\ldots,9\). We choose the coefficients for the current proposal according to~\eqref{eq:scaling}. For imitation we minimize
\(\|u_{\mathrm S}-u_{\mathrm{data}}\|^2+\eta\|u_{\mathrm{NN}}-u_{\mathrm{data}}\|^2\).
The auxiliary term gives the policy a direct gradient when the filter attenuates the gradient from the applied acceleration. We chose \(\eta=0.1\). We trained on the full dataset and on two subsets of 10,000 and 20,000 examples, each covering all training maps. For each dataset we used matched initial weights and the same number of training updates. One policy used the loss above, and the other fitted its proposal directly to the expert acceleration.

\subsection{Flight, intervention and learning}
On six unseen narrow mazes, we tested both policies with and without the layer from the same six initial states per map and three training seeds (Table~\ref{tab:paper-flights}). With the smallest training set, training with the layer gave 102 goal arrivals, compared with 60 when it was added after training. Both avoided wall contact. At larger training budgets, adding the layer after training gave more arrivals.

\begin{figure}[!t]\centering
\includegraphics[width=.70\linewidth,trim=0bp 3bp 0bp 0bp,clip]{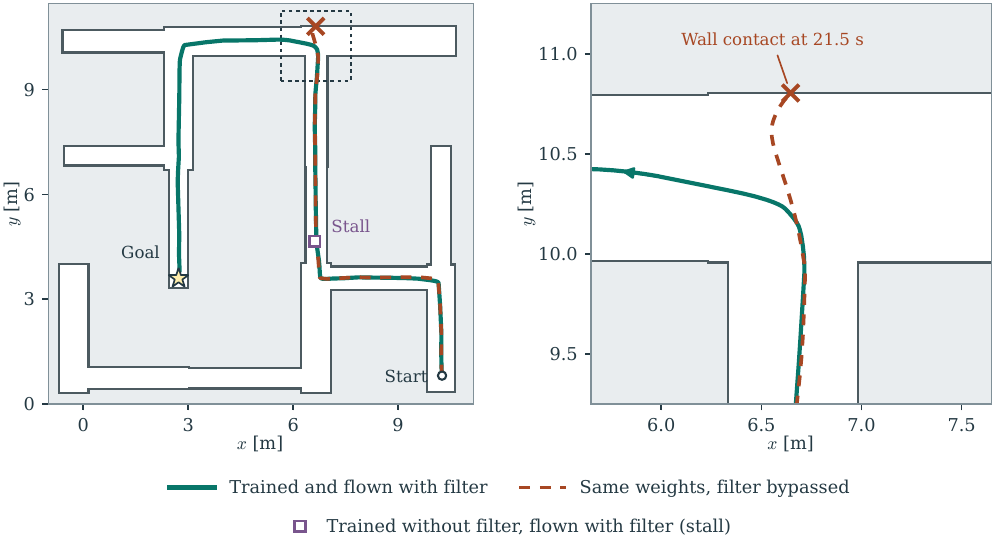}
\caption{Flight on a held-out maze. The enlargement shows the filter active and bypassed. The square marks where the drone comes to rest when the filter is added only at deployment.}\label{fig:paper-intervention}

\includegraphics[width=.76\linewidth,trim=0bp 8bp 0bp 2bp,clip]{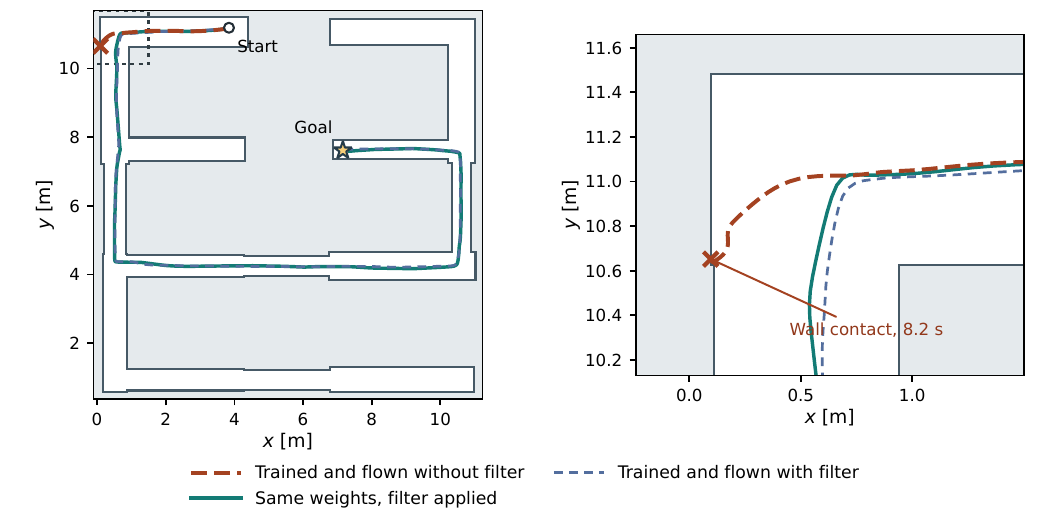}
\caption{An independently trained unfiltered policy hits the outer wall at the first turn. Applying the safety layer to the same weights allows the drone to complete the route. All three flights use fresh feedback from the same initial state.}\label{fig:raw-wall-contact}
\end{figure}

Figure~\ref{fig:paper-intervention} shows how the safety layer corrects insufficient braking before the left turn. The network still commands forward acceleration. The safety layer applies a braking acceleration, and the drone enters the next cell and reaches the goal. With the same weights and the layer bypassed, the drone carries too much speed into the turn and hits the far wall. All six contacts occurred at this bend for one training seed.

\begin{figure}[!t]\centering
\includegraphics[width=.80\linewidth,trim=0bp 1bp 0bp 0bp,clip]{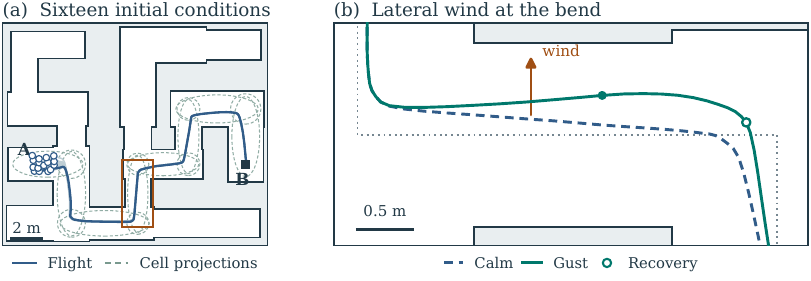}
\caption{Perturbed starts on a complex maze (left, one policy seed). The enlarged bend shows matched calm and wind-driven flights. The filled circle marks the end of the gust and the open circle marks return to the route tube.}\label{fig:paper-flight}

\includegraphics[width=.78\linewidth]{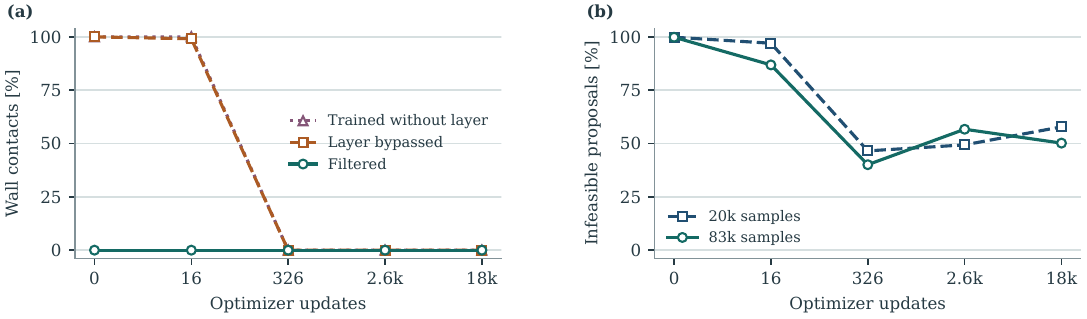}
\caption{Closed-loop evaluation of the policies in narrow test mazes at successive stages of training. Left: fraction of test flights with wall contact. Right: fraction of proposed accelerations outside the safety ellipse before filtering, for two training set sizes.}\label{fig:paper-learning}
\end{figure}

The square in Figure~\ref{fig:paper-intervention} marks where adding the layer after training brings the drone to rest. The unchanged scan and waypoint then lead to repeated zero acceleration after filtering. At this state, both policies request similar forward accelerations but different lateral components. The safety ellipse couples these components: after filtering, the policy trained with the layer retains enough forward acceleration to move. This illustrates how training with the safety layer can improve progress through a narrow bend.
On 72 further narrow mazes, the independently trained policy with the smallest training set hit a wall in 18 of 1,296 flights. Adding the layer to the same weights prevented all observed contacts. In Figure~\ref{fig:raw-wall-contact}, filtering also allowed the drone to reach the goal, although it reduced arrivals across these mazes overall.
On a broader set of familiar and unseen maps, both policies avoided wall contact and reached more goals with the full training set. Even with the full dataset, the layer corrected about two fifths of the proposals. The safety ellipse gives a sufficient condition for clearance, so a correction need not indicate an impending collision.
On a separate maze we varied the initial position and velocity and applied a lateral gust midway along the route (Figure~\ref{fig:paper-flight}). The route tube is a band extending \(0.20\,\mathrm{m}\) to either side of the planned route, used only to measure recovery. For the gust direction shown, all flights that were within this band at gust onset and subsequently left it returned with forward progress. Goal completion depended on the wind direction. Some disturbed trajectories left the nominal cell even though the applied accelerations remained feasible. The flights illustrate recovery after a disturbance. The clearance guarantee applies to the nominal motion.

We also held the first proposed acceleration for ten additional sampling intervals while continuing to update the state and safety ellipse. Both unfiltered controllers hit walls in more than half the flights. The filtered controller avoided wall contact because it could correct the held proposal using the current state at each sample.

At successive training stages, we evaluated the policies from the same initial states in the narrow test mazes, with weights fixed during each flight (Figure~\ref{fig:paper-learning}). At initialization, every unfiltered flight hit a wall, while filtering prevented contact but achieved no goals. Most early proposals were infeasible. The layer thus enforced feasibility before the policy had learned to navigate.

Finally, partially trained policies were updated during flight on two unseen layouts using MPC targets at the visited states. Roughly two thirds of the filtered policy's proposals were infeasible, yet every applied acceleration satisfied the constraint to numerical tolerance. The unfiltered policy hit walls in nearly all flights. Separate tests gave the same arrivals before and after adaptation. Thus, the layer maintained feasibility while the controller weights changed.

\subsection{Feasibility at prescribed depth and computational effort}\label{sec:computational-effort}
For safety-critical control, feasibility of the applied command, \(u_{\mathrm S}\in\mathcal C\), is required at every sampling instant, whereas proximity to the exact projection \(u^\star\) is secondary. Under the conditions of Theorem~\ref{thm:exit}, every polynomial block returns a feasible output, so the depth \(N\) can be fixed in advance without losing the feasibility guarantee. Training through the layer allows the policy to account for the correction produced by the chosen number of blocks, as illustrated by the avoided stall in Figure~\ref{fig:paper-intervention} and the limited-data comparison in Table~\ref{tab:paper-flights}.

For the polynomial layer used in the navigation example, we choose the smallest penalty weight and largest step sizes allowed by~\eqref{eq:scaling} for the current proposal. A feasible proposal then passes through unchanged. The fixed version uses an upper bound \(R\) on \(\sqrt{g(u_{\mathrm{NN}})}\), so for a given ellipse its coefficients apply to the whole prescribed range of proposals.

We first evaluated the polynomial layer on accelerations proposed by a policy trained on the full dataset, using one training seed and states visited by the MPC on 24 unseen maps. Among the roughly 740 infeasible proposals, increasing the number of blocks from one to ten reduced the mean distance to the exact projection from about \(1.4\) to \(0.1\,\mathrm{m/s^2}\). Over the same proposals, the mean of the certified bounds fell from about \(3\) to \(0.4\,\mathrm{m/s^2}\). Each bound applies to the individual command and can be evaluated without computing the exact projection. Since the recursion~\eqref{eq:certificate} does not involve the iterates, a depth sufficient for a prescribed tolerance can be determined before the blocks are evaluated.
The trained policy typically proposed accelerations that lay just outside the safety ellipse when a correction was needed. Their median normalized radius \(\sqrt{g(u_{\mathrm{NN}})}\) was about \(1.02\), with the boundary at one. For proposals farther outside the ellipse, sampled at a normalized radius of \(1.5\), the projection error decreased slowly even at 100 blocks. For a spherical constraint, the proposal-dependent choice in~\eqref{eq:scaling} with \(k_0=1\) gives the exact projection in the first block. For an elongated ellipsoid, the same rule prescribes a smaller step size as the proposal moves farther outside the safety ellipse.
With the bound implied by the full output range of the policy, the fixed-coefficient layer was too conservative at the chosen depth and all flights with this choice reached the time limit, highlighting the importance of a tight bound \(R\).

Radial scaling~\citep{rayen2023} and Frank--Wolfe iterations~\citep{lin2021fw,dfwlayer2024} also enforce feasibility, while a scalar root search~\citep{eberly2020ellipsoid} computes the Euclidean projection numerically. The polynomial iteration offers a feedforward realization whose blocks, for given coefficients, use only additions and multiplications. Its projection error is bounded in advance by~\eqref{eq:certificate}, and with the sequence of Corollary~\ref{cor:rate} its output converges to \(u^\star\) as the depth increases. Radial scaling generally returns a different feasible point. This structure makes the layer a candidate for fixed-latency hardware implementation (Section~2.1).

While MPC can be implemented efficiently even on low-end microcontrollers for small problems~\citep{Takacs2016Embedded,Nguyen2024TinyMPC,verschueren2022acados}, its computational effort generally grows with the prediction horizon and the number of constraints~\citep{diehl2009numerical,findeisen2026embedded}. For a fixed network architecture and safety-layer depth, the learned policy and its safety layer have a prescribed computational budget per command. On a single thread of an Apple M3 Max CPU, the learned policy with its safety layer took \(22\,\mu\mathrm{s}\) per command at the median. This includes feature assembly from the supplied LiDAR ranges, evaluation of the network, construction of the safety ellipse, ten polynomial steps and calculation of the error bound. On the same states, replanning and checking the 35-step MPC horizon took about \(0.8\,\mathrm{ms}\), or 36 times as long. Both medians fall within the \(100\,\mathrm{ms}\) control interval and exclude sensing and map construction. For onboard use, we also compiled an earlier version of the policy with the same architecture, including the safety layer, for an STM32 Cortex-M4F microcontroller. The controller image occupied about 36~KiB of flash and budgeted less than 5~KiB of RAM, including the stack. The timings and memory footprint support deploying the navigation policy and safety layer on resource-limited embedded systems. Sensing and global planning require a separate assessment; the planner can also run offboard.

\section{Conclusion}\label{sec:conclusion}
Training a neural network on feasible examples alone does not ensure that its outputs satisfy a hard constraint. For a learned mapping with an ellipsoidal output constraint, we constructed a safety layer that approximates the Euclidean projection of the network proposal. Each block implements one polynomial penalty step exactly for the supplied coefficients. The loss can therefore be evaluated at the filtered output and differentiated through the same blocks used at deployment. We showed how to choose the parameters of the polynomial safety layer so that every block returns a feasible output. At any chosen depth, we can also bound the distance from that output to the exact projection. The construction also applies to nonlinear discrete-time models that are affine in the control input, provided the quadratic safety condition defines a nonempty, nondegenerate ellipsoid of inputs.

We evaluated the safety layer in simulation on a drone navigating indoor mazes. We trained its LiDAR-based policy to imitate a privileged model predictive control (MPC) expert that used the map and the measured drone state to compute target accelerations. The filtered policy avoided wall contact at every tested stage, even before it had learned to navigate. During in-flight adaptation, the safety layer kept every applied acceleration feasible. On unseen mazes, adding the layer to a policy trained without filtering prevented the wall contacts observed with the same weights, although fewer flights reached the goal overall. The narrow-maze trials show that including the safety layer in training can also improve navigation performance. With limited data, the policy trained with the layer reached the goal more often than the policy trained without it, although both used the same layer in flight. This advantage disappeared with larger training sets.

We plan to extend the construction to more complex feasible sets described by several overlapping ellipsoids. In particular, we will study how to blend the outputs of their safety layers while preserving feasibility. We also plan to evaluate the policy and safety layer in closed-loop flight on a drone with limited onboard computing power, including an implementation of both as one network on an FPGA, and measure whether they meet the control deadline.
\subsection*{Generative AI use}
We used ChatGPT and Codex (OpenAI, including GPT-6 Astra) to assist with manuscript drafting and preparation of the simulation code. We carefully evaluated and checked the simulation results reported in this paper. The authors take full responsibility for the content, results, and conclusions presented in this paper.
\FloatBarrier
\label{end:body}
\bibliography{navigation_references,ten_sources,L4DC_2027_v05_refs}

@book{rawlings:mayne:diehl:2020,
  author = {Rawlings, J. B. and Mayne, D. Q. and Diehl, M. M.},
  title = {Model Predictive Control: Theory, Computation, and Design},
  edition = {2nd, paperback},
  publisher = {Nob Hill Publishing},
  address = {Santa Barbara, CA},
  year = {2020},
  isbn = {978-0-9759377-5-4},
  url = {https://sites.engineering.ucsb.edu/~jbraw/mpc/}
}

@article{findeisen03c,
  author       = {R. Findeisen and L. Imsland and F. Allg{\"o}wer and B. A. Foss},
  title        = {State and Output Feedback Nonlinear Model Predictive Control: An Overview},
  journal      = {Eur. J. Control},
  volume       = {9},
  number       = {2--3},
  pages        = {190--206},
  year         = {2003},
  doi          = {10.3166/ejc.9.190-206},
  url          = {https://doi.org/10.3166/ejc.9.190-206}
}

@inproceedings{agrawal2017dcbf,
  author       = {Agrawal, A. and Sreenath, K.},
  title        = {Discrete Control Barrier Functions for Safety-Critical Control of Discrete Systems with Application to Bipedal Robot Navigation},
  booktitle    = {Proc. Robotics: Science and Systems (RSS)},
  year         = {2017},
  address      = {Cambridge, Massachusetts},
  month        = {July},
  doi          = {10.15607/RSS.2017.XIII.073}
}

@inproceedings{Xu2024LogicDesign,
  author = {Xu, K. and Zhang, G. L. and Schlichtmann, U. and Li, B.},
  title = {Logic Design of Neural Networks for High-Throughput and Low-Power Applications},
  booktitle = {Proc. 29th Asia South Pac. Des. Autom. Conf. (ASP-DAC)},
  pages = {902--907},
  year = {2024},
  doi = {10.1109/ASP-DAC58780.2024.10473844}
}

@article{Lucia2018FPGA,
  author       = {Lucia, S. and Navarro, D. and Luc{\'i}a, {\'O}. and Zometa, P. and Findeisen, R.},
  title        = {Optimized {FPGA} Implementation of Model Predictive Control for Embedded Systems Using High-Level Synthesis Tool},
  journal      = {IEEE Trans. Ind. Informat.},
  volume       = {14},
  number       = {1},
  pages        = {137--145},
  year         = {2018},
  doi          = {10.1109/TII.2017.2719940}
}

@inproceedings{Takacs2016Embedded,
  author = {Tak{\'a}cs, G. and Zometa, P. and Findeisen, R. and Roha\v{l}-Ilkiv, B.},
  title = {Embedded Model Predictive Vibration Control on Low-End 8-Bit Microcontrollers via Automatic Code Generation},
  booktitle = {Proc. 23rd Int. Congr. Sound Vib. (ICSV)},
  address = {Athens, Greece},
  pages = {434--441},
  year = {2016},
  isbn = {978-1-5108-2716-5},
  url = {https://www.proceedings.com/content/031/031280webtoc.pdf}
}

@incollection{diehl2009numerical,
  author    = {Diehl, M. and Ferreau, H. J. and Haverbeke, N.},
  title     = {Efficient Numerical Methods for Nonlinear {MPC} and Moving Horizon Estimation},
  booktitle = {Nonlinear Model Predictive Control},
  editor    = {Magni, L. and Raimondo, D. M. and Allg{\"o}wer, F.},
  series    = {Lecture Notes in Control and Information Sciences},
  volume    = {384},
  pages     = {391--417},
  publisher = {Springer},
  address   = {Berlin, Heidelberg},
  year      = {2009},
  doi       = {10.1007/978-3-642-01094-1_32}
}

@incollection{findeisen2026embedded,
  author    = {Findeisen, R. and Rose, A. and Graichen, K. and M{\"o}nnigmann, M.},
  title     = {Embedded Optimization in Control: An Introduction, Opportunities, and Challenges},
  booktitle = {Encyclopedia of Systems and Control Engineering},
  editor    = {Ding, Z.},
  volume    = {4},
  pages     = {223--244},
  publisher = {Elsevier},
  address   = {Oxford},
  year      = {2026},
  doi       = {10.1016/B978-0-443-14081-5.00129-X}
}

@inproceedings{Nguyen2024TinyMPC,
  author    = {Nguyen, K. and Schoedel, S. and Alavilli, A. and Plancher, B. and Manchester, Z.},
  title     = {{TinyMPC}: Model-Predictive Control on Resource-Constrained Microcontrollers},
  booktitle = {Proc. IEEE Int. Conf. Robot. Autom. (ICRA)},
  year      = {2024},
  doi       = {10.1109/ICRA57147.2024.10610987}
}

@article{verschueren2022acados,
  author  = {Verschueren, R. and Frison, G. and Kouzoupis, D. and Frey, J. and van Duijkeren, N. and Zanelli, A. and Novoselnik, B. and Albin, T. and Quirynen, R. and Diehl, M.},
  title   = {{acados}---a modular open-source framework for fast embedded optimal control},
  journal = {Math. Program. Comput.},
  volume  = {14},
  number  = {1},
  pages   = {147--183},
  year    = {2022},
  doi     = {10.1007/s12532-021-00208-8}
}

@inproceedings{chen2020,
  author       = {Chen, D. and Zhou, B. and Koltun, V. and Kr{\"a}henb{\"u}hl, P.},
  title        = {Learning by Cheating},
  booktitle    = {Proc. Conf. Robot Learning (CoRL)},
  series       = {Proceedings of Machine Learning Research},
  volume       = {100},
  pages        = {66--75},
  year         = {2020},
  url          = {https://proceedings.mlr.press/v100/chen20a.html},
  publisher    = {PMLR}
}

@inproceedings{schlor2025,
  author = {Schlor, S. and Iannelli, A. and Kim, J. and Shim, H. and Allg{\"o}wer, F.},
  title = {A Polynomial-Based {QCQP} Solver for Encrypted Optimization},
  booktitle = {Proc. IEEE Conf. Decis. Control (CDC)},
  pages = {7885--7891},
  year = {2025},
  doi = {10.1109/CDC57313.2025.11312321}
}

@misc{dalal2018safe,
  title        = {Safe Exploration in Continuous Action Spaces},
  author       = {Dalal, G. and Dvijotham, K. and Vecer{\'i}k, M. and Hester, T. and Paduraru, C. and Tassa, Y.},
  year         = {2018},
  howpublished = {arXiv preprint},
  eprint       = {1801.08757},
  archiveprefix = {arXiv},
  doi          = {10.48550/arXiv.1801.08757},
  url          = {https://arxiv.org/abs/1801.08757}
}

@inproceedings{donti2021dc3,
  title        = {{DC3}: A Learning Method for Optimization with Hard Constraints},
  author       = {Donti, P. L. and Rolnick, D. and Kolter, J. Z.},
  booktitle    = {Proc. Int. Conf. Learning Representations (ICLR)},
  year         = {2021},
  url          = {https://openreview.net/forum?id=V1ZHVxJ6dSS},
  eprint       = {2104.12225},
  archiveprefix = {arXiv}
}

@inproceedings{gregor2010lista,
  title        = {Learning Fast Approximations of Sparse Coding},
  author       = {Gregor, K. and LeCun, Y.},
  booktitle    = {Proc. Int. Conf. Machine Learning (ICML)},
  pages        = {399--406},
  year         = {2010},
  url          = {https://icml.cc/2010/papers/449.pdf}
}

@article{koegel25,
  author       = {K{\"o}gel, M. and Ibrahim, M. and Kallies, C. and Findeisen, R.},
  title        = {Safe hierarchical model predictive control and planning for autonomous systems},
  year         = {2025},
  journal      = {Int. J. Robust Nonlinear Control},
  volume       = {35},
  number       = {7},
  pages        = {2658--2676},
  doi          = {10.1002/rnc.6808}
}

@article{xiao2023barriernet,
  author = {Xiao, W. and Wang, T.-H. and Hasani, R. and Chahine, M. and Amini, A. and Li, X. and Rus, D.},
  title = {{BarrierNet}: Differentiable Control Barrier Functions for Learning of Safe Robot Control},
  journal = {IEEE Trans. Robot.},
  volume = {39},
  number = {3},
  pages = {2289--2307},
  year = {2023},
  doi = {10.1109/TRO.2023.3249564}
}

@article{hsu2024safety,
  author       = {Hsu, K.-C. and Hu, H. and Fisac, J. F.},
  title        = {The Safety Filter: A Unified View of Safety-Critical Control in Autonomous Systems},
  journal      = {Annu. Rev. Control Robot. Auton. Syst.},
  volume       = {7},
  pages        = {47--72},
  year         = {2024},
  doi          = {10.1146/annurev-control-071723-102940}
}

@inproceedings{pfeiffer2017,
  author = {Pfeiffer, M. and Schaeuble, M. and Nieto, J. and Siegwart, R. and Cadena, C.},
  title = {From Perception to Decision: A Data-Driven Approach to End-to-End Motion Planning for Autonomous Ground Robots},
  booktitle = {Proc. IEEE Int. Conf. Robot. Autom. (ICRA)},
  pages = {1527--1533},
  year = {2017},
  doi = {10.1109/ICRA.2017.7989182}
}

@incollection{richter2016,
  author = {Richter, C. and Bry, A. and Roy, N.},
  title = {Polynomial Trajectory Planning for Aggressive Quadrotor Flight in Dense Indoor Environments},
  booktitle = {Robotics Research},
  editor = {Inaba, M. and Corke, P.},
  series = {Springer Tracts in Advanced Robotics},
  volume = {114},
  pages = {649--666},
  publisher = {Springer},
  address = {Cham},
  year = {2016},
  doi = {10.1007/978-3-319-28872-7_37}
}

@article{rayen2023,
  author = {Tordesillas, J. and Klemm, V. and How, J. P. and Hutter, M.},
  title = {{RAYEN}: Imposition of Hard Convex Constraints on Neural Networks},
  journal = {Int. J. Robot. Res.},
  year = {2026},
  note = {Accepted for publication},
  eprint = {2307.08336},
  archiveprefix = {arXiv},
  url = {https://arxiv.org/abs/2307.08336v3}
}

@inproceedings{lin2021fw,
  author       = {Lin, J.-L. and Hung, W. and Yang, S.-H. and Hsieh, P.-C. and Liu, X.},
  title        = {Escaping from Zero Gradient: Revisiting Action-Constrained Reinforcement Learning via {Frank-Wolfe} Policy Optimization},
  booktitle    = {Proc. Conf. Uncertainty in Artificial Intelligence (UAI)},
  series       = {Proceedings of Machine Learning Research},
  volume       = {161},
  pages        = {397--407},
  year         = {2021},
  url          = {https://proceedings.mlr.press/v161/lin21b.html},
  publisher    = {PMLR}
}

@inproceedings{dfwlayer2024,
  author       = {Liu, Z. and Liu, L. and Wang, X. and Zhao, P.},
  title        = {{DFWLayer}: Differentiable {Frank-Wolfe} Optimization Layer},
  year         = {2024},
  url          = {https://openreview.net/forum?id=XoPQqUctS1},
  booktitle    = {Tiny Papers at the Int. Conf. Learning Representations (ICLR)}
}

@inproceedings{agrawal2019cvxpylayers,
  title        = {Differentiable Convex Optimization Layers},
  author       = {Agrawal, A. and Amos, B. and Barratt, S. and Boyd, S. and Diamond, S. and Kolter, J. Z.},
  booktitle    = {Advances in Neural Information Processing Systems},
  year         = {2019},
  volume       = {32},
  publisher    = {Curran Associates, Inc.},
  url          = {https://proceedings.neurips.cc/paper_files/paper/2019/hash/9ce3c52fc54362e22053399d3181c638-Abstract.html},
  eprint       = {1910.12430},
  archiveprefix = {arXiv}
}

@inproceedings{drgona25,
  author       = {Drgo{\v n}a, J. and Nghiem, T. X. and Beckers, T. and Fazlyab, M. and Mallada, E. and Jones, C. and Vrabie, D. and Brunton, S. L. and Findeisen, R.},
  title        = {Safe Physics-Informed Machine Learning for Dynamics and Control},
  booktitle    = {Proc. Am. Control Conf. (ACC)},
  pages        = {591--606},
  year         = {2025},
  doi          = {10.23919/ACC63710.2025.11107836}
}

@article{wabersich2023filters,
  author = {Wabersich, K. P. and Taylor, A. J. and Choi, J. J. and Sreenath, K. and Tomlin, C. J. and Ames, A. D. and Zeilinger, M. N.},
  title = {Data-Driven Safety Filters: {Hamilton-Jacobi} Reachability, Control Barrier Functions, and Predictive Methods for Uncertain Systems},
  journal = {IEEE Control Syst. Mag.},
  volume = {43},
  number = {5},
  pages = {137--177},
  year = {2023},
  doi = {10.1109/MCS.2023.3291885}
}

@techreport{eberly2020ellipsoid,
  author       = {Eberly, D.},
  title        = {Distance from a Point to an Ellipse, an Ellipsoid, or a Hyperellipsoid},
  institution  = {Geometric Tools},
  year         = {2020},
  url          = {https://www.geometrictools.com/Documentation/DistancePointEllipseEllipsoid.pdf}
}
\end{document}